\pdfoutput=1

\documentclass[sigconf,screen=True,authorversion=True]{acmart}

\usepackage{algorithm}
\usepackage{algpseudocode}
\usepackage{listings}
\usepackage{enumitem}
\usepackage{float}
\usepackage{booktabs}
\usepackage{multirow}

\usepackage{comment}
\usepackage[textsize=tiny, backgroundcolor=green!50,linecolor=green!75]{todonotes}

\usepackage{caption}
\AtBeginDocument{%
  }

\copyrightyear{2024}
\acmYear{2024}
\setcopyright{rightsretained}
\acmConference[RecSys '24]{18th ACM Conference on Recommender Systems}{October 14--18, 2024}{Bari, Italy}
\acmBooktitle{18th ACM Conference on Recommender Systems (RecSys '24), October 14--18, 2024, Bari, Italy}\acmDOI{10.1145/3640457.3691707}
\acmISBN{979-8-4007-0505-2/24/10}

\usepackage[absolute]{textpos}
\begin{document}
\begin{textblock}{16}(0,0.1)\centerline{This paper was published at \textbf{RecSys 2024} -- please cite the published version {\small\url{https://doi.org/10.1145/3640457.3691707}}.}\end{textblock}

\title{beeFormer: Bridging the Gap Between Semantic and Interaction Similarity in Recommender Systems}

\author{Vojtěch Vančura}

\email{vancurv@fit.cvut.cz}
\orcid{0000-0003-2638-9969}

\affiliation{%
  \institution{Faculty of Information Technology, Czech Technical University in Prague}
  \city{Prague}
  \country{Czech Republic}
}

\affiliation{%
    \institution{Recombee}
    \city{Prague}
    \country{Czech Republic}
}

\author{Pavel Kordík}

\email{pavel.kordik@fit.cvut.cz}
\orcid{0000-0003-1433-0089}
\affiliation{%
  \institution{Faculty of Information Technology, Czech Technical University in Prague}
  \city{Prague}
  \country{Czech Republic}
}
\affiliation{%
    \institution{Recombee}
    \city{Prague}
    \country{Czech Republic}
}

\author{Milan Straka}
\email{straka@ufal.mff.cuni.cz}
\orcid{0000-0003-3295-5576}
\affiliation{%
  \institution{Faculty of Mathematics and Physics, Charles University}
  \city{Prague}
  \country{Czech Republic}
}
\begin{abstract}

  Recommender systems often use text-side information to improve their predictions, especially in cold-start or zero-shot recommendation scenarios, where traditional collaborative filtering approaches cannot be used. Many approaches to text-mining side information for recommender systems have been proposed over recent years, with sentence Transformers being the most prominent one. However, these models are trained to predict semantic similarity without utilizing interaction data with hidden patterns specific to recommender systems. In this paper, we propose beeFormer, a framework for training sentence Transformer models with interaction data. We demonstrate that our models trained with beeFormer can transfer knowledge between datasets while outperforming not only semantic similarity sentence Transformers but also traditional collaborative filtering methods. We also show that training on multiple datasets from different domains accumulates knowledge in a single model, unlocking the possibility of training universal, domain-agnostic sentence Transformer models to mine text representations for recommender systems. We release the source code, trained models, and additional details allowing replication of our experiments at {\url{https://github.com/recombee/beeformer}}.
\end{abstract}



\maketitle

\section{Introduction}

Recommender systems (RS) aim to help users find what they are looking for 
in various domains. Many approaches to building algorithms for RS have been proposed over the past years, with Collaborative Filtering (CF) \cite{tapestry} being the most popular choice. CF methods predict (filter) user preferences by analyzing past interactions. Popular CF techniques include neighborhood-based methods \cite{Konstan1997, sarwar2001item}, matrix-factorization (MF)~\cite{Koren2009,alsmf}, deep neural networks (DNN) \cite{Liang2018,cheng2016wide}, or shallow linear autoencoders (SLAs) \cite{Steck2019, sansa, ELSA}. 

\begin{figure}
    \includegraphics[angle=0,width=0.95\linewidth]{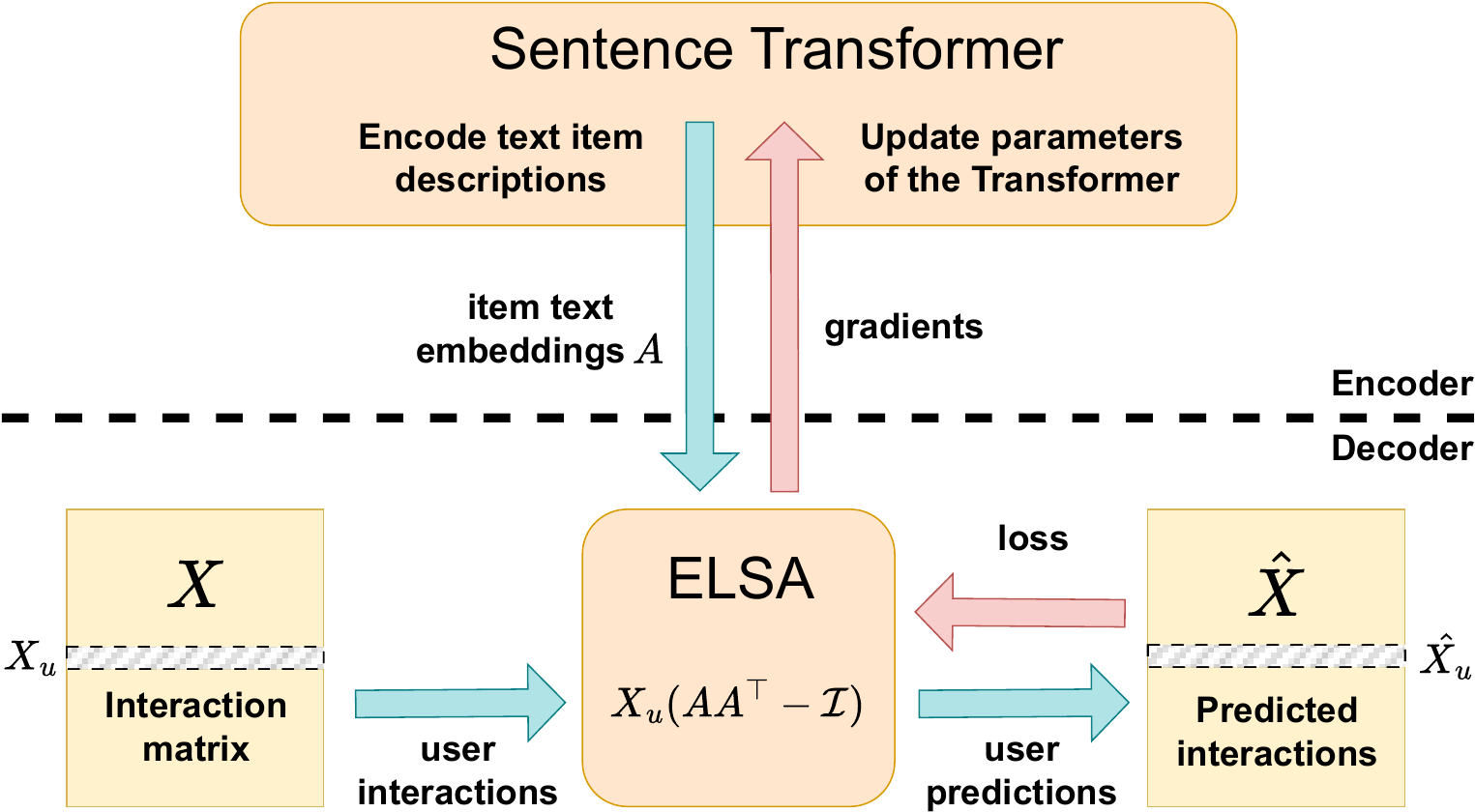}
    \caption{Training with the beeFormer framework: A sentence Transformer model (up) act as a an encoder to generate item embeddings represented by the matrix $\boldsymbol A$. Then ELSA act as a decoder (down) in a training step on interactions to obtain gradients used to optimize the Transformer model.}
    \label{fig:schema}
\end{figure}

\looseness-1
SLAs became popular recently, mainly because the EASE \cite{Steck2019} model has a closed-form solution while yielding high performance comparable to deep models. Since EASE cannot scale to datasets with a high number of items, scalable variants of EASE have been proposed: SANSA \cite{sansa} and ELSA \cite{ELSA}. SANSA keeps the original structure of EASE but uses (sparse) incomplete Cholesky factorization approximating the inverse of the (potentially) large sparse matrix $X^TX$ to construct the asymmetric approximation of the item-to-item weight matrix. ELSA, on the other hand, approximates learned item-to-item weight matrix $W$ with low-rank approximation $W = AA^T$, with $\operatorname{diag}(W)=0$ to prevent trivial solution such as the identity solution. The matrix $A$ is optimized using backpropagation.

Despite %
the popularity and state-of-the-art performance  of CF methods in recommendation tasks, they cannot provide any predictions when there are no interactions. In such cases, also known as 
cold-start \cite{sethi2021cold} and zero-shot \cite{ding2022zero} recommendation, one can use content-based filtering (CBF) ~\citep{baeza2015predicting,anwaar2018hrs} using side information (attributes, images, text) directly to produce recommendations or learn a transformation function to transform side information to CF representations \cite{zhu2020recommendation,du2020learn}.

Using texts (item descriptions, user reviews, etc.) as side information has become extremely popular after the invention of the Transformer \cite{vaswani2017attention} neural architecture. Transformer models can be used to encode text into a vector representation, and sentence Transformers \cite{reimers-2019-sentence-bert} were explicitly developed to mine text latent representations from the whole blocks of text (
sentences, paragraphs), which can then be used for various tasks, with the recommendation being one of them.

However, sentence Transformers trained to predict semantic similarity often fail to capture patterns and user behaviors hidden in the interaction data. In many cases, users may look for a specific item (for example, batteries when buying a kid's toy, or cables when buying a new printer) with very low semantic similarity compared to other items in the catalog.

\looseness-1
To bridge this gap between the semantic and the interaction similarity, we employ the following idea: We use the training procedure from the ELSA model, but instead of optimizing the matrix $A$, we generate matrix $A$ with a sentence Transformer model
and optimize parameters of the sentence Transformer instead of optimizing $A$ directly, as illustrated in Figure~\ref{fig:schema}. However, this approach faces one critical problem: In every training step, we need to generate and optimize embeddings for all items in the catalog, which leads to very high effective batch size for Transformer training, e.g., possibly over a million for some datasets. We propose to overcome this problem by employing the following three techniques: gradient checkpointing~\cite{gradcheck}, gradient accumulation~\cite{chen2016training_accugrad}, and negative sampling~\cite{ding2020simplify}. Combining these techniques with sentence Transformer and ELSA training procedure, we present beeFormer (short for the Recom\textbf{bee} Trans\textbf{former} in camel case), a sentence Transformer training framework that uses text-side information and interactions directly to update the parameters of a Transformer model.

The main contributions of this paper are listed as follows:

\begin{itemize}[topsep=2pt, itemsep=0pt,leftmargin=0.8cm]
    \item We propose beeFormer, a framework for training sentence Transformers on interaction data with text-side information.
    \item Our experiments show that sentence Transformer models trained with beeFormer outperform all baselines in cold-start, zero-shot and time-split recommendation scenarios.
    \item We demonstrate the beeFormer's ability to transfer knowledge between datasets.
    \item We show that training models on combined datasets from various domains further increase performance in the domain-agnostic recommendation.
    \item We create and publish LLM-generated item descriptions for all used datasets for reproducibility of our experiments. 
    \item Models trained with beeFormer are easily deployable into production systems using the sentence Transformers library. 
\end{itemize}
\noindent We believe the above improvements open a path towards a potentially universal, domain-agnostic, and multi-modal Transformer models for recommender systems. %

\section{Related Work}

SLAs \cite{ Steck2019, sansa, ELSA} have recently gained much attention, mainly because of their simplicity while retaining performance comparable to deep models, with EASE \cite{Steck2019} being the most promising. ELSA~\cite{ELSA} solved the scalability issue of EASE and enabled its use on large datasets by low-rank approximation of the item-to-item weight matrix of EASE. We use the idea of training the ELSA's embeddings using backpropagation to obtain the gradients for training the sentence Transformer model.

The first attempt to use Transformers in RS was in sequential recommendation. Bert4rec \cite{BERT4Rec} used item IDs as tokens and treated the sequential recommendation problem in the same manner as NLP. Several improvements to this approach have been proposed since then \cite{transformers4rec, shalaby2022m2trec, P5}. Promising approach is to create artificial text sequences combining descriptions of interacted items and train a Transformer-based model on them \cite{text_is_all_you_need, universal_text_seq}. 

Another direction is to use Large Language Models (LLMs), such as the Chat-GPT \cite{di2023evaluating_chatgpt_rec,dietmar_gpt}. %
LLMs can be used for various tasks, e.g., as conversational recommendation \cite{sun2024llm_convrec_study, conv_rec_adapting}, to generate standardized item text descriptions \cite{llm_item_descriptions}, recommendation explanation \cite{chat_gpt_alan_martjen}, or to produce recommendations through its text output \cite{llm_cf} directly. %

Sentence Transformers \cite{reimers-2019-sentence-bert} use a pooling function on top of the Transformer architecture and provide a robust, easy-to-use framework for tokenization, embedding generation, and training sentence Transformer models. While using sentence Transformers is very popular in the dense-retrieval domain \cite{botha-etal-2020-entity, gillick-etal-2019-learning}, using sentence Transformers in recommender systems is limited to generating side information for cold-start methods \cite{zhu2020recommendation,du2020learn} or using neural networks \cite{sentence_neural} or graph neural networks \cite{sentence_graph} on top of sentence Transformer models.

Improving sentence Transformers with training on interaction data is crucial for boosting the performance of all methods using sentence embeddings as side information mentioned above. Sadly, to our best knowledge, there is no prior work on training sentence Transformers directly with interaction data in the RS domain.

\section{Training Procedure}
We follow notation from \cite{ELSA}: assume a set of users $\mathbb{U} = \{u_1,u_2,\ldots, \\u_{U}\}$, a set of items $\mathbb{I} = \{i_1,i_2,\ldots,i_{I}\}$, and a set of token sequences representing corresponding items $\mathbb{T} = \{t_1,t_2,\ldots,t_{I}\}$ . Let $X \in \{0,1\}^{|\mathbb U| \times |\mathbb I|}$ be a user-item interaction matrix:  $X_{a,b}=1$ if the user $u_a$ interacted with item $i_b$, and  $X_{a,b}=0$ otherwise. Assume that $M_{a}$ is a \emph{column} vector corresponding to the $a$-$th$ row of matrix $M$. Let $\operatorname{norm}(M)$ be a function that 
$L^2$-normalizes each row of a matrix, so $\operatorname{norm}(M)_a = M_a / \|M_a\|$.
Finally, let $g(\bullet, \theta_g) : \mathbb{T}  \xrightarrow{} \mathbb{R}^{I\times d}$ be a Transformer-based neural network with parameters $\theta_g$.

The training procedure starts with generating latent representations of items -- the matrix $A$:
            \begin{equation}\label{A_forward}
                A = g(\mathbb{T}, \theta_g). \\
            \end{equation}
Then, we can compute our loss \cite{ELSA} as:
            \begin{equation}\label{A_forward}
                L = \big\| \operatorname{norm}\big( X_u \big)- \operatorname{norm}\big( X_u (AA^\top - \mathcal{I}) \big)\big\|^2_{F}. \quad   \\%\text{s.t.} \quad  ||A_{a,.}||^2 = 1 \\
            \end{equation}

Finally, we compute the gradients of $L$ with respect to $A$, and then the gradients for $\theta_g$ are computed using the chain rule. However, when using %
common deep learning frameworks, there is a practical problem regarding the memory needed to track gradients for $\theta_g$ because the size of matrix $A$ depends on the number of \textit{all} items.

We employ the following procedure (described in Algorithm
~\ref{alg:cap}) to address this memory problem: First, we compute the matrix $A$ in batches without tracking %
the gradients for $\theta_g$ (lines 1-5 of the algorithm). Then, we compute predictions (line 8), loss (lines 10-13) for a batch of users $X_u$, and \textit{gradient checkpoint}~\cite{gradcheck} for %
the matrix $A$ (lines 14,16). Finally, we compute %
the matrix $A$ in batches again and use %
the gradient checkpoint to compute %
the gradients for $\theta_g$. We \textit{accumulate gradients}~\cite{chen2016training_accugrad} during the loop (lines 19-21). Finally, we update $\theta_g$ with a PyTorch optimizer (line 22).

\begin{algorithm}
\caption{beeFormer training step procedure in Python using PyTorch}\label{alg:cap}
\begin{flushleft}
 \textbf{Input:} \\
 \hspace{\algorithmicindent}\text{batch of interactions \lstinline[columns=fixed]{X}}\\
 \hspace{\algorithmicindent}\text{Transformer model \lstinline[columns=fixed]{transformer}}\\
 \hspace{\algorithmicindent}\text{sequences of tokens \lstinline[columns=fixed]{tokenized_texts}}\\
 \hspace{\algorithmicindent}\text{PyTorch optimizer \lstinline[columns=fixed]{optimizer} optimizing  the parameters} \\ 
 \hspace{\algorithmicindent}\text{of the Transformer model \lstinline[columns=fixed]{transformer}}\\ 
\textbf{Output:} \\
 \hspace{\algorithmicindent}\text{Transformer model with updated weights  \lstinline[columns=fixed]{transformer}} \\
 \hspace{\algorithmicindent}\text{predictions  \lstinline[columns=fixed]{X_pred}} \\
 \hspace{\algorithmicindent}\text{computed loss \lstinline[columns=fixed]{loss}}
\\\hrulefill

\begin{lstlisting}[language=Python,morekeywords={with,True},numbers=left,xleftmargin=0.7cm]
with torch.no_grad():
    A_list = []
    for t in tokenized_texts:
        A_list.append(transformer(t))
    A = torch.vstack(A_list)

A.requires_grad = True
X_pred = X @ A @ A.T - X

loss = torch.nn.MSELoss(
    torch.nn.functional.normalize(X),
    torch.nn.functional.normalize(X_pred),
)
loss.backward()

checkpoint = A.grad

optimizer.zero_grad()
for i, t in enumerate(tokenized_texts):
    A_i = transformer(t)
    A_i.backward(gradient=checkpoint[i])
optimizer.step()
\end{lstlisting}
\end{flushleft}
\end{algorithm}
Using the algorithm above, we effectively enable %
training of %
a Transformer model using gradients computed with the ELSA algorithm on top of it%
, from the memory point of view. However, computing $A$ for all items in every training step quickly becomes time-consuming for datasets with large number of items. At this point, we would like to note one property specific to recommender systems: interaction matrix $X$ is typically (very) sparse. This property means that when we sample a random batch of user interaction vectors from $X$, we obtain interactions only with a limited number of items, meaning that it is not necessary to encode all items to matrix $A$ in every training step. More formally, let $\mathbb{I}_b$ be a subset of $\mathbb{I}$ derived from the interactions presented in a random batch $b$ sampled from $X$. Recent work shows that significant improvement in performance during training CF models can be achieved by negative sampling~\cite{ding2020simplify}. We implement negative sampling by adding random items to $\mathbb{I}_b$ and fixing the total number of items observed in each step during training to $m$. The $m$ becomes an additional hyperparameter fulfilling $|\mathbb{I}_b|\le m\ll|\mathbb I|$ for any possible $\mathbb{I}_b$, to control the size of the matrix $A$. Finally, in a situation when $|\mathbb{I}_b|<m$, we %
select uniformly at random $m-|\mathbb{I}_b|$ %
items from the catalog and add them to the sampled batch from $X$ with zeros in corresponding columns. A~notable advantage of our approach is that the asymptotic complexity of beeFormer does not depends on the overall number of items $|\mathcal I|$ but only on the hyperparameter $m$.
 
\section{Experimental Evaluation}

We %
evaluate our models on several popular datasets for evaluating recommender systems: \textbf{MovieLens20M} (ML20M) \cite{movieLens}, \textbf{Goodbooks-10k}  (GB10k)~\cite{goodbooks2017}%
, and \textbf{Amazon Books} (AB)~\cite{amazon_data}. Since these datasets contain explicit ratings, we transform them into implicit feedback datasets by considering a rating of four or higher as an interaction between the user and the item, and we keep only users with at least five interactions. Then, we collect movie plots from the IMDB Movies Analysis dataset~\cite{Mhatre_2020_imdb} to obtain the descriptions for the items of the MovieLens20M dataset, and we collect book descriptions from the Goodreads books - 31 features dataset~\cite{Reese_2020_goodreads} and Goodreads 100K books dataset~\cite{Dhamani_2021_goodreads_100k} to get the descriptions for the Goodbooks-10k dataset; most items in the Amazon Books dataset already contains descriptions. We then remove the items without descriptions.
Finally, we %
use the Meta-Llama-3.1-8B-Instruct\footnote{LLAMA 3.1 license allows the use of generated output to train new language models. We add the prefix "Llama" to the names of our models to comply with license terms.}~\cite{llama3} to generate standardized item descriptions from the existing ones to train our models.\footnote{Using Llama to generate the item descriptions allows us to publish them. More details about the item description generation are available on our GitHub page.} Since %
the LLM %
refuse to generate descriptions for some items (for example, because it refuses to generate explicit content), we %
remove such items from the dataset. %
 Summary of %
the resulting dataset's properties is available in Table~\ref{tab:datasets}.

\begin{table}[]
\caption{Detailed statistics of datasets used for evaluation.}
\label{tab:datasets}
\setlength{\tabcolsep}{10pt}
\begin{tabular}{lrrr}
\toprule
                                 & GB10k    & ML20M    & AB      \\ \midrule
\# of items in $X$                 & 9 975    & 16 902   & 63 305  \\
\# of users in $X$                 & 53 365   & 136 589  & 634 964 \\
\# of interactions in $X$          & 4.20 M   & 9.69 M   & 8.29 M  \\
density of $X$                     & 0.77 \%  & 0.42 \%  & 0.02 \% \\
density of $X^TX$ & 41.22 \% & 26.93 \% & 7.59 \% \\ \bottomrule
\end{tabular}
\end{table}

We %
use two setups for our experiments. First, to test the ability of beeFormer to generalize toward new items, we split ML20M and GB10k datasets item-wise (item-split): we randomly %
choose %
2000 items as the test set, and we %
use the rest to cross-validate and train our models. Next, we simulate a real-world scenario with the AB dataset: we sorted all interactions by timestamp and used the last 20\% of interactions as a test set (time-split). Again, the remaining interactions were used as validation and training sets.

\subsubsection*{Item-split Setup}
In the item-split setup, we use the following scenarios: %
Firs, a zero-shot scenario, where the models were not trained on the evaluated dataset -- they %
need to transfer knowledge from other datasets. We %
use CBF in this scenario -- we %
generate item embeddings with a sentence Transformer, and then we %
use cosine similarities between item embeddings to provide recommendations. %
Second, in a cold-start scenario%
,
the models %
use %
the text side information to generalize towards new, previously unseen items. We %
use the Heater %
model~\cite{zhu2020recommendation} %
mapping interaction data to side information, to %
benchmark the baseline models. Again, for our models%
, we %
use CBF.

We choose to compare %
our beeFormer-trained models to three %
best-performing sentence Transformer models:
\begin{itemize}[topsep=2pt, itemsep=0pt,leftmargin=0.6cm]
    \item \textbf{all-mpnet-base-v2}, the best performing model from the sentence Transformers \cite{reimers-2019-sentence-bert} library. It is based on MPNET \cite{song2020mpnet}, a model pre-trained with combined masked and permuted language modeling and then finetuned on various datasets.
    \item \textbf{BAAI/bge-m3} \cite{chen2024bge}, which uses three-fold versatility (dense, sparse/lexical, and multi-vector) retrieval during training. It is trained and finetuned on various datasets and synthetic data.
    \item \textbf{nomic-ai/nomic-embed-text-v1.5} \cite{nussbaum2024nomic}, which uses Flash attention \cite{dao2022flashattention} to handle longer context (up to 8192 tokens for predictions.) It was trained using a Contrastor framework.\footnote{\url{https://github.com/nomic-ai/contrastors}}
\end{itemize}

\subsubsection*{Time-split Setup}
Similarly, for the time-split setup, we use a zero-shot scenario and %
supervised %
models trained with interaction data. This setup allows the use of classical CF; we chose KNN~\cite{sarwar2001item}, ALS matrix factorization~\cite{takacs_implict}, ELSA~\cite{ELSA}, and SANSA~\cite{sansa} as baselines. %

We %
train four %
models by finetuning the initial \textit{all-mpnet-base-v2} model: one for each dataset and one model combining %
the data from %
the ML20M and GB10k datasets (denoted %
\emph{goodlens}). The resulting models are available on our Huggingface page.%
\footnote{\url{https://huggingface.co/beeformer}}

We use Recall@20 (R@20), Recall@50 (R@50), and NDCG@100 (N@100) metrics computed using the RecPack \citep{recpack} framework to compare the models. We also calculate the standard error for each experiment with bootstrap resampling.

We publicly share further details about the datasets, description of the LLM generative procedure, resulting data for reproducibility, hyperparameters, and other technical details %
on our GitHub page
, along with all the source code.\footnote{\url{https://github.com/recombee/beeformer}}

\subsection{Results}

\subsubsection*{Item-split setup} In the zero-shot scenario, we observe that %
the beeFormer models trained on a different domain (books vs. movies) significantly outperform all baselines. Detailed results are in Table~\ref{tab:results-zero}.

\begin{table}[]
\caption{Results for zero-shot scenario in item-split setup. {\normalfont Names of our models trained with beeFormer are in italics, the best-performing models are represented in bold, and the best baseline for each scenario is underlined. The standard error of all values is below 0.0001 (evaluated via bootstrap resampling).}}
\label{tab:results-zero}
\setlength{\tabcolsep}{4.5pt}
\begin{tabular}{@{}lllll@{}}
\toprule
Dataset                & Sentence Transformer             & \multicolumn{1}{r}{R@20} & \multicolumn{1}{r}{R@50} & \multicolumn{1}{r}{N@100} \\ \midrule
\multirow{4}{*}{GB10K} & all-mpnet-base-v2                & 0.1017                   & 0.1886                   & 0.1739                    \\
                       & nomic-embed-text-v1.5            & {\underline{0.1146}}             & {\underline{0.2069}}             & {\underline{0.1896}}              \\
                       & bge-m3                           & 0.1134                   & 0.1953                   & 0.1838                    \\
                       & \textit{Llama-movielens-mpnet} & 0.1782          & 0.2837          & 0.2719           \\ 
                       & \textit{Llama-amazbooks-mpnet} & \textbf{0.2649}          & \textbf{0.3957}          & \textbf{0.3787}           \\ 
                       \midrule
\multirow{4}{*}{ML20M} & all-mpnet-base-v2                & 0.0788                   & 0.1550                   & 0.1042                    \\
                       & nomic-embed-text-v1.5            & 0.1113                   & {\underline{0.2143}}             & 0.1511                    \\
                       & bge-m3                           & {\underline{0.1409}}             & 0.2125                   & {\underline{ 0.1578}}              \\
                       & \textit{Llama-goodbooks-mpnet} & \textbf{0.1589}          & \textbf{0.2647}          & \textbf{0.2066}           \\ \bottomrule
\end{tabular}
\end{table}

For the cold-start scenario, we observe that beeFormer-trained models outperform all baselines when the Heater model approach is used for mapping semantic embeddings to interactions. The %
model %
demonstrate interesting behavior when trained on multiple datasets: %
\emph{goodlens} model outperforms the models trained solely on the evaluated dataset both for ML20M and GB10k. This indicates the possibility to train one (universal) recommender model on multiple datasets from multiple domains. Detailed results are in Table~\ref{tab:results-cold}.

Comparing results from Tables \ref{tab:results-zero} and \ref{tab:results-cold}, models trained on different datasets within the same domain yield similar performance to models trained on the evaluated dataset. This demonstrates the critical capability of beeFormer to transfer knowledge from one dataset to another. 

\subsubsection*{Time-split setup} We %
utilize the time-split %
setup to compare %
the beeFormer-trained models with %
the CF models. The beeFormer models %
outperform all CF baselines for both training on the evaluated dataset and %
in the zero-shot scenario within the same domain. The model trained on multiple datasets performs slightly worse in cold-start scenario than pure in-domain knowledge transfer, but the results are still comparable. Detailed results are in Table~\ref{tab:time-split}.

\begin{table}[]
\caption{Results for cold-start scenario in item-split setup. {\normalfont Names of our models trained with beeFormer are in italics, the best-performing models are represented in bold, and the best baseline for each scenario is underlined. The standard error of all values is below 0.0001 (evaluated via bootstrap resampling).}}
\label{tab:results-cold}
\begin{tabular}{@{}lllll@{}}
\toprule
\begin{tabular}[c]{@{}l@{}}Dataset\\ Method\end{tabular}                 & Sentence Transformer             & R@20         & R@50         & N@100        \\ \midrule
\multirow{2}{*}{\begin{tabular}[c]{@{}l@{}}GB10K\\ CBF\end{tabular}}     & \textit{Llama-goodbooks-mpnet} & 0.2505       & 0.3839       & 0.3747       \\
 & \textit{Llama-goodlens-mpnet}  & \textbf{0.2710} & \textbf{0.4218} & \textbf{0.4066} \\ \cmidrule(l){2-5} 
\multirow{4}{*}{\begin{tabular}[c]{@{}l@{}}GB10K\\ Heater\end{tabular}}  & all-mpnet-base-v2                & 0.2078       & 0.3221       & {\underline{0.3195}} \\
 & nomic-embed-text-v1.5            & {\underline{0.2154}}    & {\underline{0.3317}}    & 0.3193          \\
 & bge-m3                           & 0.2052          & 0.3113          & 0.3099          \\
 & \textit{Llama-movielens-mpnet} & 0.2060          & 0.3161          & 0.3196          \\ \midrule
\multirow{2}{*}{\begin{tabular}[c]{@{}l@{}}ML20M\\ CBF\end{tabular}}    & \textit{Llama-movielens-mpnet} & 0.4291       & 0.6108       & 0.4054       \\
 & \textit{Llama-goodlens-mpnet}  & \textbf{0.4630} & \textbf{0.6152} & \textbf{0.4066} \\ \cmidrule(l){2-5} 
\multirow{4}{*}{\begin{tabular}[c]{@{}l@{}}ML20M\\ Heater\end{tabular}} & all-mpnet-base-v2                & {\underline{0.3114}} & {\underline{ 0.4331}} & {\underline{ 0.3407}} \\
 & nomic-embed-text-v1.5            & 0.3049          & 0.4285          & 0.3270          \\
 & bge-m3                           & 0.2847          & 0.3932          & 0.3161          \\
 & \textit{Llama-goodbooks-mpnet} & 0.3204          & 0.4669          & 0.3381          \\ \bottomrule
\end{tabular}
\vspace*{-8pt}
\end{table}

\begin{table}[]
\caption{Results for time-split setup on the Amazon Books dataset. {\normalfont Names of our models trained with beeFormer are in italics, the best-performing models are represented in bold, and the best baseline for each scenario is underlined. The standard error of all values is below 0.00005  (evaluated via bootstrap resampling).}}
\label{tab:time-split}
\setlength{\tabcolsep}{3.9pt}
\begin{tabular}{@{}lllll@{}}
\toprule
\begin{tabular}[c]{@{}l@{}}Scenario\\ Method\end{tabular} & Model                            & R@20            & R@50            & N@100           \\ \midrule
\multirow{6}{*}{\begin{tabular}[c]{@{}l@{}} zero-shot\\CBF\end{tabular}} & all-mpnet-base-v2 & 0.0218 & 0.0336 & 0.0193 \\
                                                          & nomic-embed-text-v1.5            & 0.0387          & {\underline{0.0560}}    & {\underline{0.0320}}    \\
                                                          & bge-m3                           & {\underline{0.0398}}    & 0.0546          & 0.0313          \\
                                                          & \textit{Llama-goodbooks-mpnet} & \textbf{0.0649} & \textbf{0.0931} & \textbf{0.0515} \\
                                                          & \textit{Llama-goodlens-mpnet}  & 0.0617          & 0.0891          & 0.0492          \\ \midrule
\multirow{3}{*}{\begin{tabular}[c]{@{}l@{}} supervised\\CF\end{tabular}}                                       
                                                          & KNN                              & 0.0370          & 0.0562          & 0.0303          \\
                                                          & ALS MF                           & 0.0344          & 0.0580          & 0.0313          \\
                                                          & ELSA                             & 0.0367          & 0.0628          & 0.0346          \\
                                                          & SANSA                            & {\underline{0.0421}}    & {\underline{0.0678}}    & {\underline{0.0362}}    \\
CBF                                                       & \textit{Llama-amazbooks-mpnet} & \textbf{0.0706} & \textbf{0.1045} & \textbf{0.0571} \\ \bottomrule
\end{tabular}
\vspace*{-8pt}
\end{table}

\section{Conclusions}

We introduce beeFormer, a novel training procedure that enhances neural representations of items by training sentence encoders on interactions. Our approach is scalable, utilizing ELSA linear autoencoder as the decoder during the training process, enabling it to handle %
datasets with a large number of items. BeeFormer-trained models are easily deployable into existing production systems since they are compatible with the widely adopted sentence Transformers library and can produce recommendations via embedding tables with cosine similarity criterion.

We observe performance improvements over various state-of-the-art baselines in all evaluated scenarios. Notably, in the time-split setup on the Amazon Books dataset, our models %
achieves significantly better results over %
CF methods in both supervised and zero-shot scenarios, demonstrating both superior performance and the ability to transfer knowledge from one dataset to another. %

We also %
demonstrate that training the \emph{Llama-goodlens-mpnet} model on two datasets (GB10K and ML20M) from different domains further increases performance when evaluating on individual datasets. This ability to accumulate knowledge from multiple datasets marks an important step towards training universal, domain-agnostic, content-based models for RS. 

In our future work, we %
plan to build one (big) dataset from several RS domains and train a universal %
sentence Transformer model on it. We %
want to also explore the possibility of using beeFormer with computer vision models. Building multi-modal encoders with beeFormer could be especially useful in %
domains such as fashion recommendation.

\begin{acks}
We want to thank anonymous reviewers for their suggestions, many of which helped us improve this paper.
Our research has been supported by the Grant Agency of Czech Technical University (SGS23/210/OHK3/3T/18) and by the Grant Agency of the Czech Republic under the EXPRO program as project ``LUSyD'' (project No.~GX20-16819X).
\end{acks}

\bibliographystyle{ACM-Reference-Format}
\bibliography{sample-base}

\end{document}